\pdfoutput = 1

\documentclass[doublecol]{epl2}

\title{Complex Networks Analysis of Language Complexity}
\shorttitle{Network Analysis of Language Complexity} 

\author{Diego R. Amancio\inst{1} \and Sandra M. Aluisio\inst{2} \and Osvaldo N. Oliveira Jr.\inst{1} \and Luciano da F. Costa\inst{1}}
\shortauthor{D. R. Amancio \etal}

\institute{
  \inst{1} Institute of Physics of S\~ao Carlos \\
	University of S\~ao Paulo, P. O. Box 369, Postal Code 13560-970 \\
	S\~ao Carlos, S\~ao Paulo, Brazil \\ \ \\
  \inst{2}
  Institute of Mathematics and Computer Science \\
	University of S\~ao Paulo, P. O. Box 369, Postal Code 13560-970 \\
	S\~ao Carlos, S\~ao Paulo, Brazil \\
}


\usepackage{color}
\usepackage{multirow}
\definecolor{greencolor}{rgb}{0,0.5,0.2}
\definecolor{redcolor}{rgb}{1.0,0.,0.}
\definecolor{bluecolor}{rgb}{0,0.,1.}
\definecolor{greycolor}{rgb}{.5,.5,.5}

\usepackage[nearskip,caption=false,font=footnotesize]{subfig}
\usepackage[pdftex]{hyperref}

\abstract
{Methods from statistical physics, such as those involving complex networks, have been increasingly used in quantitative analysis of linguistic phenomena. In this paper, we represented pieces of text with different levels of simplification in co-occurrence networks and found that topological regularity correlated negatively with textual complexity. Furthermore, in less complex texts the distance between concepts, represented as nodes, tended to decrease. The complex networks metrics were treated with multivariate pattern recognition techniques, which allowed us to distinguish between original texts and their simplified versions. For each original text, two simplified versions were generated manually with increasing number of simplification operations. As expected, distinction was easier for the strongly simplified versions, where the most relevant metrics were node strength, shortest paths and diversity. Also, the discrimination of complex texts was improved with higher hierarchical network metrics, thus pointing to the usefulness of considering wider contexts around the concepts. Though the accuracy rate in the distinction was not as high as in methods using deep linguistic knowledge, the complex network approach is still useful for a rapid screening of texts whenever assessing complexity is essential to guarantee accessibility to readers with limited reading ability.}

\begin{document}

\maketitle

\section{Introduction}

Statistical physics has been applied in the analysis of a variety of phenomena from social sciences and linguistics~\cite{cross,liuXu,ortuno,carpena}, in many cases permitting unprecedented interpretation based on quantitative measurements~\cite{science1,exploring}.
Of particular importance for the present study has been the use of complex networks in treating linguistic issues~\cite{cross,liuXu,name,evolution2,lnet}, including those associated with natural language processing (NLP) tasks~\cite{sum,r1,r2,author}. One normally exploits the finding that networks deriving from text exhibit a scale-free topology, where the degree {distribution} follows a power law, regarded as a consequence of the rich-get-richer paradigm~\cite{beyond}. Examples of application of network concepts in NLP include strategies for automatic summarization~\cite{sum}, evaluation of machine translations~\cite{r1,r2}, analysis of lexical resources~\cite{name}, language evolution~\cite{evolution2} and authorship recognition~\cite{author}.

In this paper, we combine metrics extracted from networks representing text and pattern recognition methods to investigate complexity in texts. Motivation for this endeavor came from the need to assess whether written material in the Internet is accessible to widespread communities, including people with low levels of education. This is especially important for countries such as Brazil, for which official figures revealed that in 2009 7\% of the population were classified as illiterate; 21\% as literate at the rudimentary level; 47\% as literate at the basic level; and only 25\% as literate at the advanced level\footnote{\href{http://www.ipm.org.br}{http://www.ipm.org.br}}. Concerted efforts have been made to develop methods to detect and simplify complex textual structures, with the aim of making information accessible to people who are not able to read complex texts. Although the implementation of simplification strategies is the most important item in applications requiring text simplification, before applying any technique one should first identify the pieces of texts considered complex. Most importantly, one has to unveil the characteristics that make a text difficult to read. Indeed, many approaches have been developed to quantify textual complexity\footnote{Some approaches are presented in the Supplementary Information (SI), available from \href{https://dl.dropbox.com/u/2740286/supplementary.pdf}{https://dl.dropbox.com/u/2740286/supplementary.pdf}}~\cite{sida}, but there is still no consensus on how complexity can be measured effectively. Here we propose a new approach based on a possible correlation between textual complexity and regularity of network topology. More specifically, we apply the methodology described in Refs.~\cite{r1,r2,author}, which combines topological characterization based on descriptors with pattern recognition strategies, to evaluate complexity in texts with distinct levels of simplification. We shall show that considering wider contexts around words in the text is useful for the intended complexity discrimination.

\section{Methodology}

Networks are employed to represent texts, whose topology is examined through several metrics. The patterns emerging from the topological features are investigated with clustering and supervised learning techniques in order to correlate with textual complexity.

\subsection{Database}

The database was developed under the PorSimples project, available online from \url{http://www2.nilc.icmc.usp.br/wiki/index.php/English}. All $113$ texts were collected from the Brazilian Zero Hora newspaper~\footnote{http://www.zerohora.com.br} and their simplified versions were created by a linguist expert in textual simplification. For each original (non-simplified) text there are two corresponding simplified versions, which differ from each other by the number of operations applied to simplify the text (see the list of possible operations in {Table S2 of the SI}). The first one, referred to as natural simplification, was obtained with only a few simplification operations. In the version obtained with the procedure referred
 to as strong simplification, all the possible simplification operations were performed. The statistics related to the three corpora and examples of original texts and simplifications are given in {Figure S1 and in Table S3 of the SI}, respectively.

\subsection{Network Formation}
\begin{figure}
    \begin{center}
        \includegraphics[width=0.49\textwidth]{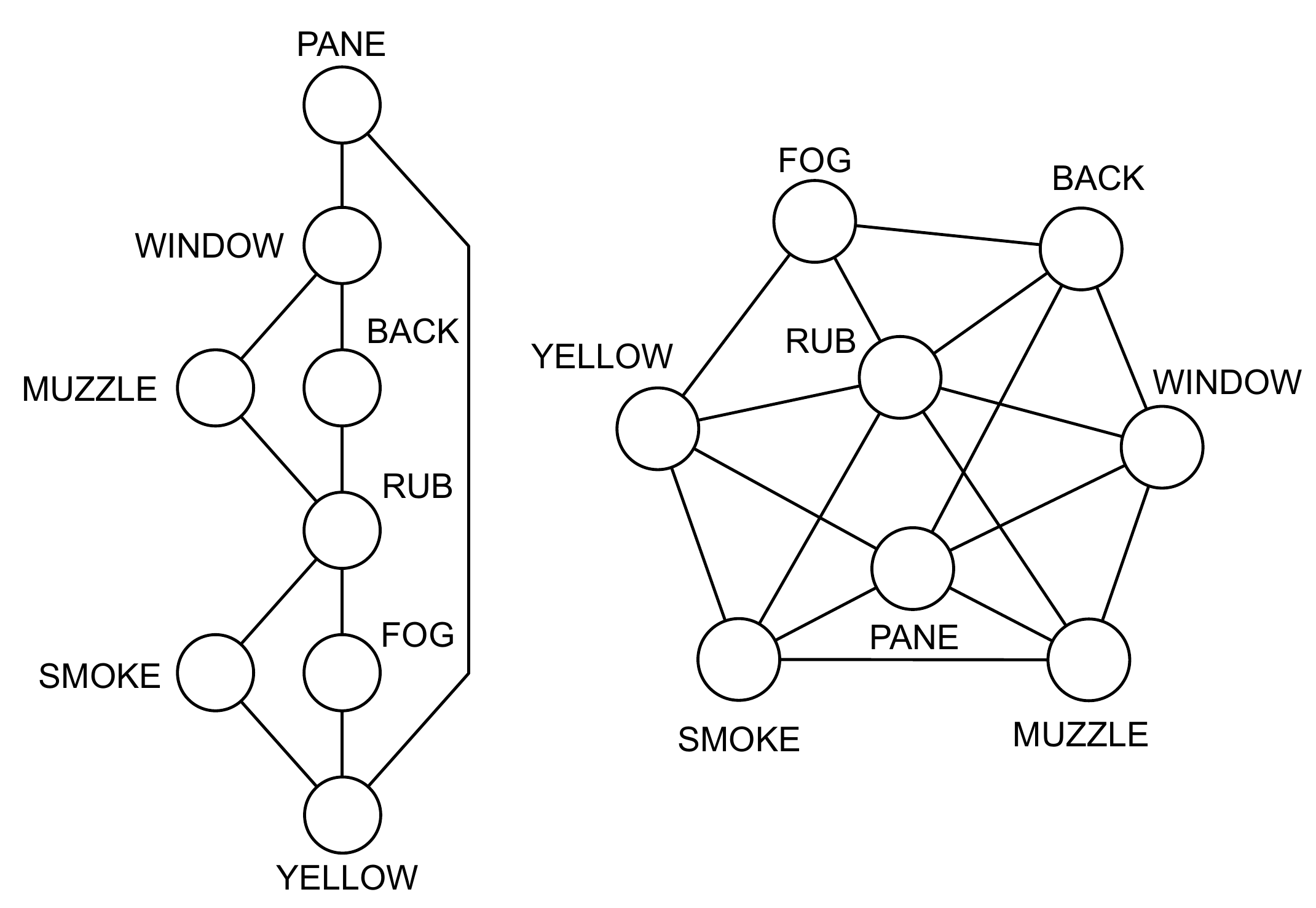}
    \end{center}
    \caption{\label{topologia} Network for the sentences ``The yellow fog that rubs its back upon the window-panes // The yellow smoke that rubs its muzzles on the window-panes'', extracted from ``The Love Song of J. Alfred Prufrockby'', by T.S. Elliot. The text from the pre-processing step is: ``yellow fog rub back window pane yellow smoke rub muzzle window pane''. Note that stopwords were removed from the original text and the remaining words were lemmatized. The network depicted on the left side illustrates the model that links words whenever they appear immediately adjacent in the pre-processed text (i.e. $w=1$). In the network illustrated on the right side, two words were connected if there was at most one intermediate word separating them in the pre-processed text (i.e. $w=2$). }
    \label{fig.6}
\end{figure}

To model texts as complex networks, preprocessing steps were applied. {\it Stopwords}\footnote{{\it Stopwords} are very frequent words usually conveying little semantic meaning, such as articles and prepositions.} were eliminated and the remaining words were lemmatized. That is to say, words were converted to their singular (nouns) and infinitive forms (verbs), so that words with different inflections but related to the same concept were taken as a single node in the network. To perform this conversion, ambiguities were resolved by using the MXPost part-of-speech tagger based on the Ratnaparki's model~\cite{ratim}. Then, each word in the pre-processed text was represented as a node and edges were established depending on the distance separating words in the preprocessed text. If words $i$ and $j$ were separated by less than $w$ intermediate words, then a connection $i \rightarrow j$ was established. To take into account the repetition of word associations, edges were weighted by the number of times the association appeared in the preprocessed text. This process is illustrated in Figure \ref{fig.6}, where a subgraph obtained for a small text extract is shown for $w = 1$ and $w = 2$.

\subsection{Complex network metrics}

The measurements to characterize complex networks can be local or global, describing a node or the whole network, respectively. The global measurements were computed using the first two statistical moments of the distribution: $\langle \mu \rangle = n^{-1} \sum \mu(i)$ and $\Delta \mu = n^{-1} \sum ( \mu(i) - \langle \mu \rangle ) ^ 2$, where $n$ is the number of nodes and $\mu(i)$ is the value of metric $\mu$ computed for node $i$. These statistical moments were chosen because they have been useful for characterizing complete networks through the analysis of local measurements~\cite{author,evolution2}. The metrics to characterize texts were: strength ($s$), clustering coefficient ($C$), average shortest path length \footnote{The computation of the average shortest paths length only makes sense in networks whose edge weights represent the distance between nodes. Since edge weights in text networks are given by the frequency of associations, correlating to similarity, we had to introduce a relationship between dissimilarity and distance. This was done by transforming the edge weights $\omega$, as follows. $l_a$ is the average shortest path length computed when the transformation $\omega \rightarrow 1 / \omega$ is performed. $l_b$ is the analogous for the transformation $\omega \rightarrow \omega_{max} - \omega$, where $\omega_{max}$ is the maximum value assumed by $\omega$ considering all edges.} ($l$), betweenness ($B$), diversity ($\delta$) and locality index ($\iota$). The hierarchical level of these metrics was computed, with the second and third levels being denoted respectively by $\mu_2$ and $\mu_3$. 

\subsection{Clustering nodes of complex networks}

Machine learning techniques are useful to recognize complex patterns and make intelligent decisions based on training examples (or input).
In this paper machine learning algorithms were used to distinguish topological factors of words and texts. In the unsupervised approach, the objective was to cluster similar instances, which in our case were words considered similar whenever their complex network measurements took similar values. The algorithm was based on the technique described in Ref~\cite{seeking}, where the attribute space was formed with metrics from complex networks. The dataset obtained with the attributes were projected onto a two-dimensional space using Principal Component Analysis~\cite{duda}. This reduction in dimensionality avoided problems stemming from high dimensional datasets, and could eliminate possible trivial correlations between the metrics. The distribution of the two-dimensional data was then estimated by the non-parametric Parzen window technique~\cite{duda}, using  Gaussians as kernel functions. Finally, each cluster was associated with a peak of the probability density function. Further details concerning this algorithm are provided in the SI.

\subsection{Supervised pattern recognition methods}

Supervised learning aims at inferring a function from the training data, whose classes are known beforehand. It may be used, for instance, in recognizing patterns and classifying unknown objects. Here, it was used to distinguish between complex and simplified texts. The simplest method is the $k$ nearest neighborhood~\cite{duda} ($k$NN), which classifies an unknown instance  with a voting process over the $k$ nearest instances. Thus, the prevailing class among the $k$ nearest instances is chosen to classify an unknown instance. In the experiments the parameter $k$ was set to $k = \{1,2,3\}$. The other machine learning algorithms used were: C4.5~\cite{duda}, which generates a decision tree as structure of decision, RIPPER~\cite{duda}, which generates a set of rules as structure of decision, and the Naive Bayes~\cite{duda} based on a probabilistic paradigm. Details regarding these algorithms are given in the SI.

\section{Results and Discussion}

The relationship between topology of complex networks and textual complexity was studied according to three different perspectives. First we investigate how local topological measurements vary with the simplification process. Then, we extended the approach to identify groups of words with the same topological feature in the network, drawing a parallel between topological regularity and textual complexity. Finally, we studied the problem from a multivariate point of view to distinguish complex texts from their simplified versions.

\subsection{Comparing original texts and their simplified versions}

We investigated changes in network topology induced by simplification by computing the global measurements for {directed, weighted networks}. For each local measurement $\mu$, we computed the relative difference $d\mu = ( \mu_s - \mu_o ) / \mu_o $ between the values for the simplified ($\mu_s$) and original texts ($\mu_o$). Significantly, we found that the average out-strength $\langle s_{out} \rangle$ tends to increase when a text is simplified, while the geodesic distances $\langle l_a \rangle$  and $\langle l_b \rangle$ are decreased ({see Figure S4 of the SI}).

\begin{figure}
			\begin{center}
            \subfloat[]
			 {\includegraphics[width=0.24\textwidth]{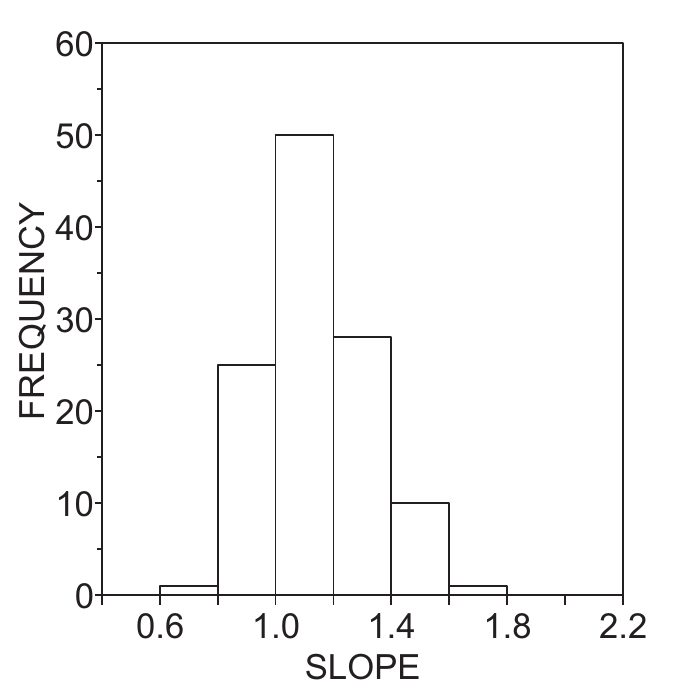}}
			\subfloat[]
			 {\includegraphics[width=0.24\textwidth]{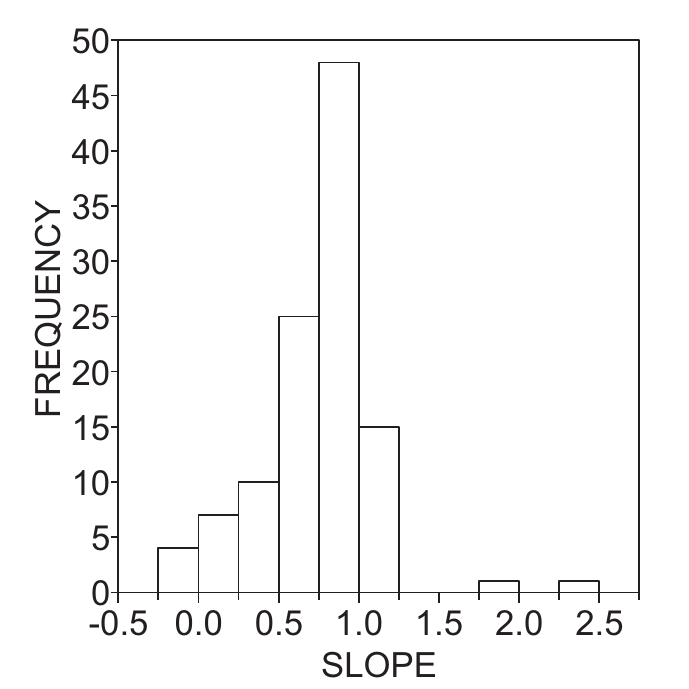}}
            \\
            \subfloat[]
			 {\includegraphics[width=0.24\textwidth]{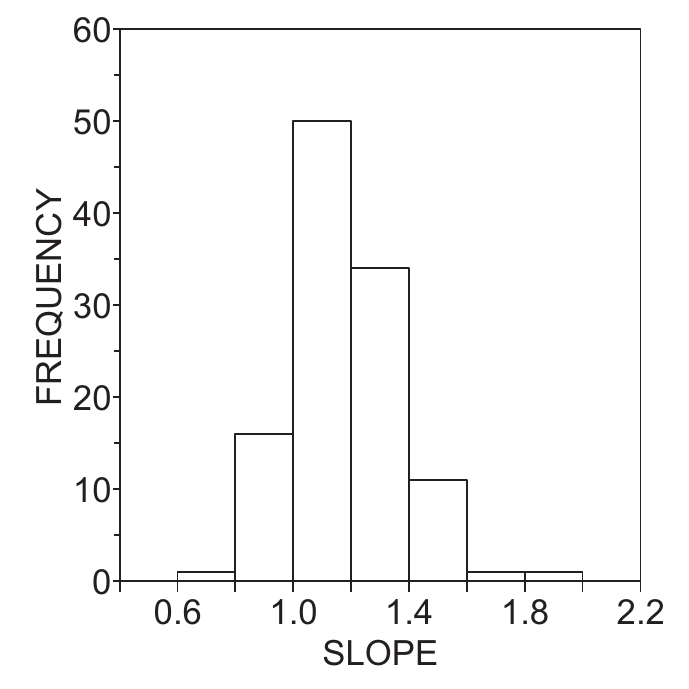}}
            \subfloat[]
			 {\includegraphics[width=0.24\textwidth]{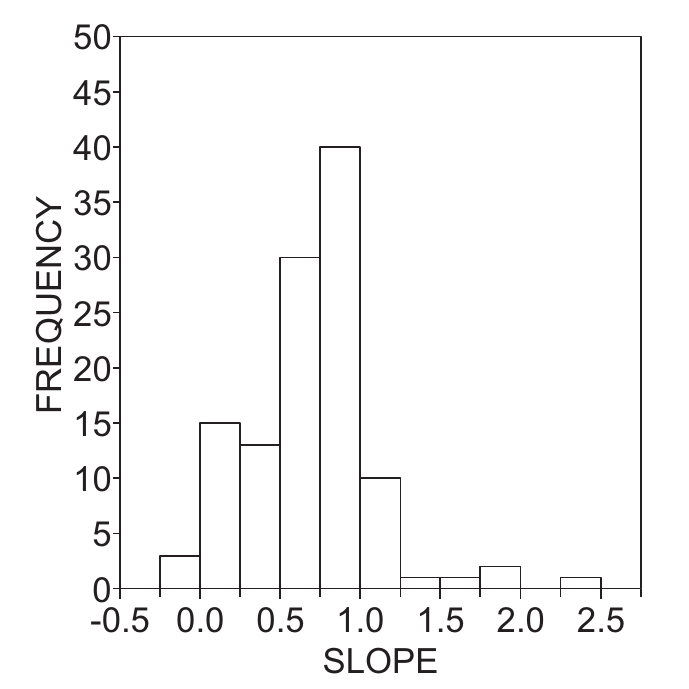}}
            \end{center}
			\caption[\it ]{Histogram of slope for (a) $s$ (natural simplification); (b) $l_a$ (natural simplification); (c) $s$ (strong simplification); and (d) $l_a$ (strong simplification). While the local $s$ tends to increase as texts are increasingly simplified, $l_a$ tends to decrease, so that distances between concepts are shortened.}
			\label{fig3}
\end{figure}

The local changes in network topology were analyzed by plotting the local measurements extracted from the original and the corresponding simplified text, node by node, so each distinct measurement leads to a scatter plot. In other words, for a given metric $\mu$ of node $v$, $\mu^{o}_v$ is the value of $\mu$ for $v$ in the original network and $\mu^{s}_v$ is $\mu$ for the same node in the simplified network, with the point ($\mu^{o}_v$, $\mu^{s}_v$) belonging to the scatter plot. Then, two descriptors were extracted from each scatter plot: the slope $m$ and the Pearson product-moment correlation coefficient $r$, obtained from the best straight line that can fit the data. These descriptors are important to provide information about possible topology preservation~\cite{r1}. For example, if a measurement tends to be preserved, the scatter plots will approach a straight line with both coefficients close to 1 {(see example in Figure S5 of the SI)}. Upon analyzing the distributions of these coefficients, we could identify a pattern similar to that in the analysis of global metrics. For $s$ and $l_a$, the correlation distributed around $1$ (results not shown). Therefore, the distribution of points in the scatter plot approximates a straight line. With regard to the slope, $s$ showed a distribution predominantly above 1 while $l_a$ displayed a histogram with values below 1 predominating, as shown in Figure \ref{fig3}. Therefore, simplified texts tend to have increased local strength. Consequently, the shortest paths become shorter.

The patterns indicate that increasingly simplified texts tend to possess more interconnected concepts. Moreover, in simplified texts, it is easier to move from one concept to another. These findings are consistent with the expectation that in a simplified text the same words should appear more often, while fewer intermediate words are used to maintain a simple, concise vocabulary for understanding.

\subsection{Correlation between simplicity and regularity}

We may draw a parallel between text complexity and complexity of networks by investigating how the network metrics vary upon simplification. We examine the hypothesis that simplified texts induce a network structure that is also simple, so that the topology around the nodes is stable. The change in each network metric was calculated through the coefficient of variation $c_v$ defined by the ratio of standard deviation to the mean. 
Basically, $c_v$ is a normalized measure of dispersion of a probability distribution that gives high values for irregular distributions. Firstly, we performed an analysis for each single measurement. We identified the metrics able to distinguish simplifications and original texts according to the coefficient of variation of the measurement. For each pair of text (original - simplified) $c_v$ was computed to check the fraction of texts in which there is a pattern revealing that $c_v$ is larger in the original or simplified texts. The percentage of cases in which there was a predominance of larger $c_v$ values for a given class of texts are shown in Table \ref{tab.4}. Note that in $6$ out of $9$ cases, the patterns of simplified texts are more regular, since $c_v$ was larger in the original texts. Also, the diversity at the second level $\delta_2$, the strength at the third level $s_3$ and the clustering coefficient $C$ are particularly important since they have a high tendency to be stable in simplified texts, especially when strong simplifications are considered.


\begin{table*}
	\centering
    \caption{\label{tab.4}
        Percentage of pairs of texts whose coefficient of variation is larger for the original texts than for the natural/strong simplifications. In six out of nine measurements the networks for the simplified texts are more regular. The $p$-value corresponds to the likelihood of the observed percentage to be obtained if the patterns were random. Most Reg. indicates the type of text (original or simplified with the most regular structure).
    }
		\begin{tabular}{|c|c|c|c|c|c|}
			\hline
			$\mu$ & \textbf{Natural} & {\bf $p$-value} & \textbf{Strong} & {\bf $p$-value} & {\bf Most Reg.} \\
		\hline
		$\delta$        & 53.98 \% & $2.3 \times 10^{-1}$ & 61.94 \% & $7.0 \times 10^{-3}$  & Simplified \\
		$\delta_2$      & 75.22 \% & $3.7 \times 10^{-8}$ & 80.53 \% & $1.8 \times 10^{-11}$  & Simplified \\
		$\delta_3$      & 73.45 \% & $3.0 \times 10^{-7}$ & 77.88 \% & $1.0 \times 10^{-9}$ & Simplified \\
		$s$             & 14.16 \% & $1.2 \times 10^{-15}$ & 7.96 \%  & $< 1.0 \times 10^{-15}$ & Original \\
		$s_2$           & 79.64 \% & $7.4 \times 10^{-11}$ & 86.72 \% &  $< 1.0 \times 10^{-15}$ & Simplified \\
		$C$             & 79.64 \% & $7.4 \times 10^{-11}$ & 81.41 \% & $4.4 \times 10^{-12}$ & Simplified \\
        $C_2$           & 13.27 \% & $< 1.0 \times 10^{-15}$ & 16.81 \% & $2.0 \times 10^{-13}$& Original \\
        $C_3$           & 73.45 \% & $3.0 \times 10^{-7}$  & 71.68 \% &  $2.3 \times 10^{-6}$ & Simplified \\
		$\iota$         & 10.62 \% & $< 1.0 \times 10^{-15}$ & 3.64 \% & $< 1.0 \times 10^{-15}$ & Original \\
		\hline
		\end{tabular}

\end{table*}


The analysis of topological regularity was also performed by adopting a multivariate approach. Each node of the network was characterized using all the measurements described in the methodology. Then the clusters (i.e., groups of similar nodes) emerging from this characterization were detected with the clustering algorithm based on the Parzen-window density estimation (details are provided in the SI). The number of clusters in each text was taken into account because this quantity is valuable to quantify the regularity of the networks. For example, when a network encompasses many clusters, then its nodes have different characteristics. In contrast, if the network is formed by a few clusters, then there exist only a few patterns of connectivity, thus leading to a more regular topology. Following this reasoning, we tested the hypothesis that simplified texts tend to generate fewer clusters, since we assume that they are more regular. To compare the number of clusters obtained in the networks, two methods were employed. In the first, the absolute number of clusters $\zeta_{\rm abs}$  was taken. In the second one, we took the normalized number of clusters $\zeta_{\rm norm} = ({\zeta_{\rm abs} - \overline{\zeta}_{\rm rand}}) / {\sigma_{\zeta_{\rm rand}}}$,
where $\zeta_{\rm abs}$ was normalized by the mean $\overline{\zeta}_{\rm rand}$ and standard deviation $\sigma_{\zeta_{\rm rand}}$ of the number of clusters obtained in $20$ equivalent random networks ($z$-score normalization). This normalization is performed to quantify if the absolute number of clusters found is greater than that expected just by chance.

The distribution of the relative difference for the absolute and normalized number of clusters between original and simplified texts
showed that $\zeta_{\rm abs} > 0$ in 72~\% of the texts ($p$-value = $2.3 \times 10^{-6}$) and  $\zeta_{\rm norm} > 0$ in 73 \% ($p$-value = $9.0 \times 10^{-7}$) {(see Figure S6 of the SI)}. Therefore, original texts tend to have nodes agglomerated in a way that the number of clusters is larger than in the simplified texts. In other words, simplified texts tend to generate less patterns of connectivity, suggesting that linguistic simplicity is reflected in topological simplicity. {Note that similar patterns appeared for both $\zeta_{\rm abs}$ and $\zeta_{\rm norm}$, which confirms that our analysis was not affected by the size of the networks.}

\subsection{Using pattern recognition techniques to characterize complexity} \label{using}

\begin{figure}[h]
			\begin{center}
            \subfloat[]{\includegraphics[width=0.24\textwidth]{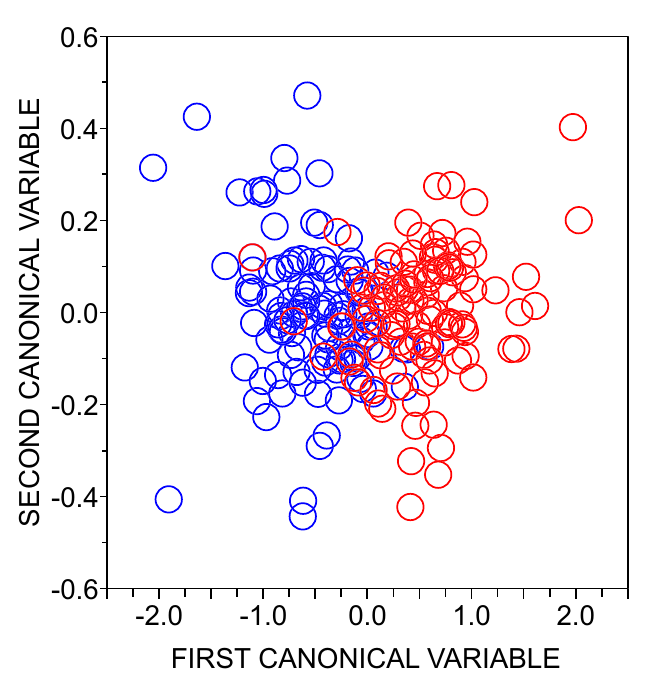}}
			 \subfloat[]{\includegraphics[width=0.24\textwidth]{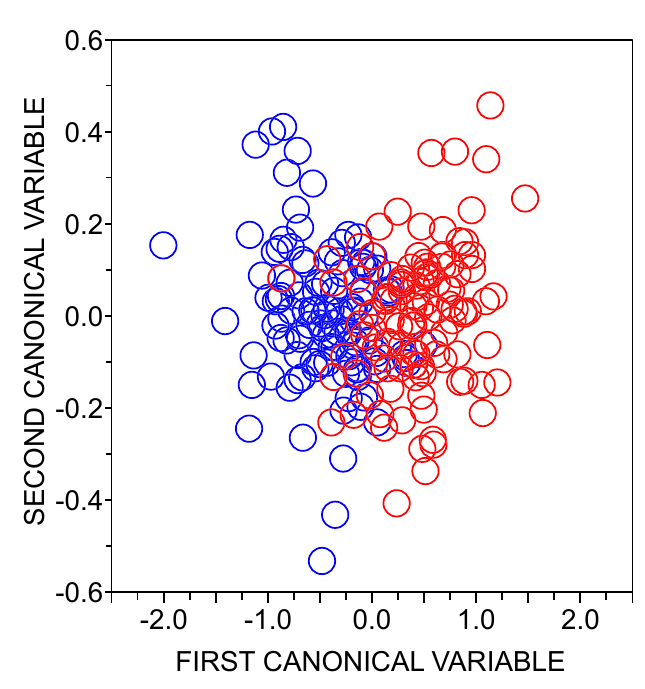}} \\
			 \subfloat[]{\includegraphics[width=0.24\textwidth]{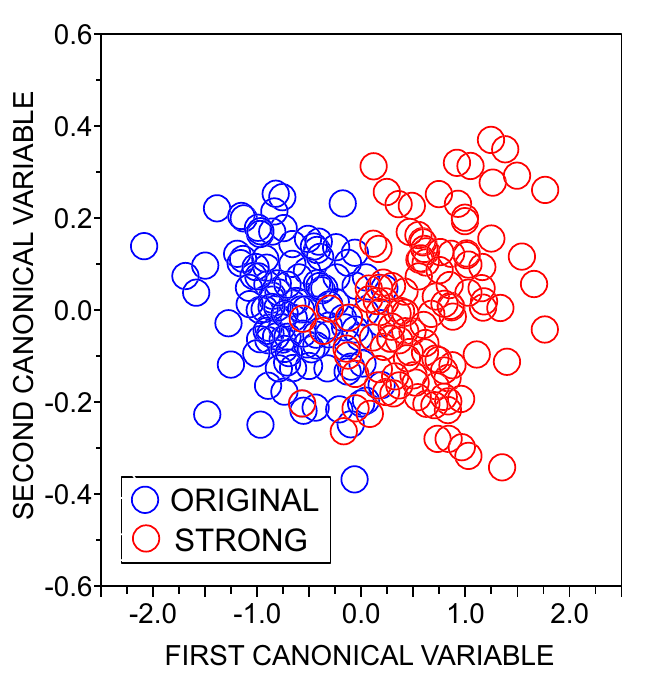}}
            \subfloat[]{\includegraphics[width=0.24\textwidth]{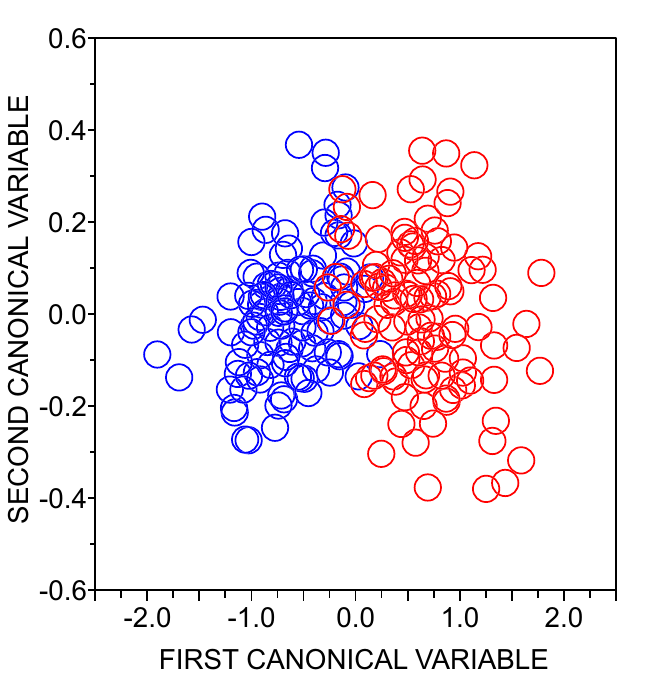}}
            \end{center}
			\caption[\it ]{Canonical Variable Analysis~\cite{duda} for original texts and strong simplifications with (a) $w=1$; (b) $w=2$; (c) $w=5$; (d) $w=6$. Better distinction was achieved when deeper levels were taken into account. The results for $w=3$ and $w=4$ are shown in Figure S8 of the SI.}
			\label{fig5}
\end{figure}

	In order to distinguish between original and simplified texts, all the global measurements were computed for each of the $339$ texts of the database. To form the networks, we linked neighbors whose distance ranged from $w=1$ to $w=6$, as explained in the methodology. The data extracted from the original and simplified texts were projected to identify the different classes. The results are shown in Figure \ref{fig5} for the original and strong simplifications {(see an analogous result for original and natural simplifications in {Figure S7 of the Supplementary Information (SI))}}. {Interestingly, the first canonical variable is sufficient to explain the separation between original and strongly simplified texts. Thus, it is possible to discriminate texts according to their complexity using a simple linear combination of network metrics.}
The ability to separate the two types of text was found to be higher for strong simplifications, and increases with the size of the window (number of intermediate words or neighborhood) for building the networks. Therefore, immediate neighborhoods, which are widely used in modeling texts as networks, are not able to fully capture distant relationships that characterize complexity. It appears that complexity might be related to the distance between concepts.

The ability to distinguish simplified from original texts was quantified using all the topological measurements as attributes to describe the texts, rather than the two canonical variables used in the projections above. This was done with the following pattern recognition techniques: C4.5, RIPPER, Naive Bayes and $k$NN (with $k=\{1,2,3\}$). They were trained to generate classifiers {(see an example of an induced classifier in Figure S9 of the SI)} whose results were evaluated using the 10-fold cross-validation technique~\cite{kohavi}. Tables \ref{tab.ref1} and \ref{tab.11} bring the results for original vs. strong simplifications and for original vs. natural simplifications, respectively. The $p$-value represents the likelihood of obtaining the corresponding accuracy rate in a random classification. Consistent with the projections, it is easier to identify the strongly simplified texts. Again, the best results for each inductor do not occur at the nearest neighborhood. We compared these results with the accuracy rate from more sophisticated networks models, where links were established according to the relations of syntax dependency~\cite{sintaxe}. The results are provided in {Tables S4 and S5 of the SI}. Interestingly, the additional effort to perform the parsing is not worthwhile, since the accuracy rates for the syntactic models are lower than those found for the model based on neighborhood. Hence, for this type of application, syntactic networks~\cite{sintaxe} were not useful to characterize textual complexity.

\begin{table}[h]
		\centering
		\caption{ \label{tab.ref1}
        Accuracy rate for the pairs of original texts and strong simplifications. All the classifications are statistically significant, which suggests a strong correlation between the complexity of texts and the topological metrics of complex networks.
        }
        \begin{tabular}{|c|c|c|c|}
			\hline
            $w$ & Accuracy & $p$-value & Classifier  \\
            \hline
            1 & 82.30~\% & $2.0 \times 10^{-13}$ & RIPPER \\ 
            2 & 84.96~\% & $7.3 \times 10^{-15}$ & RIPPER \\ 
            3 & 84.07~\% & $4.0 \times 10^{-14}$ & RIPPER \\ 
            4 & 85.40~\% & $3.5 \times 10^{-15}$ & kNN-3  \\ 
            5 & 86.28~\% & $1.2 \times 10^{-15}$ & Bayes  \\ 
            6 & 85.40~\% & $3.5 \times 10^{-15}$ & Bayes  \\ 
            \hline 		
        \end{tabular}

\end{table}

\begin{table}[h]
 \caption{ \label{tab.11}
        Accuracy rate for the pairs of original texts and natural simplifications. All the classifications are statistically significant, pointing to a strong correlation between the complexity of texts and the topological metrics.
        }		
\centering
        \begin{tabular}{|c|c|c|c|}
        \hline
        $w$ & Accuracy & $p$-value & Classifier  \\
        \hline
        1 & 71.81~\% & $2.3 \times 10^{-6}$ & RIPPER \\
        2 & 73.12~\% & $3.0 \times 10^{-7}$ & RIPPER \\
        3 & 75.33~\% & $3.6 \times 10^{-8}$ & C4.5   \\
        4 & 71.81~\% & $2.3 \times 10^{-6}$ & Bayes  \\
        5 & 72.68~\% & $9.0 \times 10^{-7}$ & Bayes  \\
        6 & 74.89~\% & $3.8 \times 10^{-8}$ & RIPPER \\
        \hline
        \end{tabular}

\end{table}

In order to compare the ability of identifying simplified texts in our approach with methods based on deep linguistic knowledge, we characterized each text of the corpus with $15$ linguistic features according to Ref.~\cite{sanda}. As one should expect in cases where further linguistic knowledge is available, a higher accuracy was obtained in comparison with the network analysis performed here ({see Section $6$ of the SI}). Nevertheless, the difference in performance is approximately  10 \%. Therefore, while the use of complex network metrics \emph{alone} is not the best approach to identify complexity, it is still useful when linguistic information is not available. Also, the strategy devised here could be combined with linguistic features to enhance the power of discrimination.

\section{Conclusion} \label{conclusao}

In this paper we described methods to evaluate textual complexity, with original and simplified texts being distinguished based on the metrics of the complex networks they represented. Important metrics for such distinction were strength, shortest paths, diversity and hierarchical measurements. The best classification achieved an accuracy of 75 \% for natural simplifications and 86 \% for strong simplifications. Moreover, we found that the measurements extracted from complex networks can be successfully applied
to classify texts
. We believe that the methodology developed here would be especially suitable for applications related to textual simplification, where the detection of complex structures is essential for obtaining concise texts.

Perhaps the main implication from the results in this paper is not related to the accuracy rates of the classifiers induced by the pattern recognition techniques. Actually, the most important results are related to the emergence of two specific patterns in complex texts: long distances between concepts and heterogeneous patterns characterized by a large number of clusters in the network. Both patterns are important because they are probably correlated with cognitive factors which make a text complex, such as the amount of information that must be memorized in the mental processes~\cite{ighnve}.

With regard to further work to be performed, one may propose a general platform to assess textual complexity. This platform may consist basically of the combination of machine learning methods and various other complex networks measurements. {We also intend to use complementary CN tools, such as the framework employed to analyze the interplay between network topology and function in complex physiological systems~\cite{ivanov}.} Finally, concepts from information theory and pattern recognition methods based on complex networks may be combined to characterize complexity in terms of the effort required to resolve semantic and phonologic ambiguities. In fact, the results presented here can be taken as evidence of the suitability of complex networks for this type of analysis, which can now be employed in applications where decisions depend on text complexity.
	


\acknowledgments
The authors would like to acknowledge CNPq (Brazil) and FAPESP (2010/00927-9 and 2011/50761-2) for the sponsorship.

\end{document}